\documentstyle[prb,aps,epsf,floats]{revtex}

\begin{document}
\renewcommand{\textfraction}{0.0}
\renewcommand{\floatpagefraction}{.7}
\setcounter{topnumber}{5}
\renewcommand{\topfraction}{1.0}
\setcounter{bottomnumber}{5}
\renewcommand{\bottomfraction}{1.0}
\setcounter{totalnumber}{5}
\setcounter{dbltopnumber}{2}
\renewcommand{\dbltopfraction}{0.9}
\renewcommand{\dblfloatpagefraction}{.7}

\draft


\title{Fixed points of Hopfield type neural networks}

\author{Leonid B. Litinskii}
\address{High Pressure Physics Institute of Russian Academy of Sciences\\
Russia, 142092 Troitsk Moscow region, e-mail: litin@hppi.troitsk.ru}


\maketitle

\begin{abstract}
The set of the fixed points of the Hopfield type network is under
investigation. The connection matrix of the network is constructed according
the Hebb rule from the set of memorized patterns which are treated as 
distorted copies of the standard-vector. It is found that the dependence of 
the set of the fixed points on the value of the distortion parameter can be 
described analytically. The obtained results are interpreted in the terms of 
neural networks and the Ising model.
\end{abstract}

${\bf 1^\circ.}$ The problem of maximization of a symmetric form which is 
quadratic in 
{\it spin} variables $\sigma_i$:
$$\left\{\begin{array}{l} F(\vec\sigma)=\sum_{i,j=1}^n J_{ij}\sigma_i\sigma_j
\to {\rm max},\ \sigma_i=\{\pm 1\},\\
\vec\sigma=(\sigma_1,\ldots,\sigma_n),
\ J_{ij}=J_{ji},\ i,j=1,2,\ldots,n.
\end{array}\right.\eqno(1)$$
is under investigation. This problem arises in the Ising model, in the 
surface physics, in the theory of optimal coding, in factor analysis, in the 
theory of neural networks and in the optimization theory 
\cite{Hopf,Dom,Surf1,Sour,Par,Brav2,Lit1}. 
Here the aim is to obtain an 
effective method for the search of the global maximum of the functional and 
a constructive description of the set of its local extrema.

The $n$-dimensional vectors $\vec\sigma$, which define $2^n$ configurations,
will be called {\it the configuration vectors}. The configuration vector which
gives the solution of the problem (1) will be called {\it the ground state}.
We investigate the problem (1) in the case of the
connection matrix constructed with regard to the Hebb rule from 
the $(p\times n)$-matrix $\bf S$  of the form
$${\bf S}=\left(\begin{array}{ccccccc}
1-x&1&\ldots&1&1&\ldots&1\\
1&1-x&\ldots&1&1&\ldots&1\\
\vdots&\vdots&\ddots&\vdots&\vdots&\ldots&\vdots\\
1&1&\ldots&1-x&1&\ldots&1\end{array}\right), \eqno(2)$$
where $x$ is an arbitrary real number. We introduce the special notation 
${\bf N}$ for the related connection matrix:
$${\bf N=S^T\cdot S},\quad N_{ii}=0,\quad i=1,2,\ldots,n.\eqno(3)$$

According to the conventional neural network tradition we treat the
$n$-dimensional vectors $\vec s^{(l)}$, which are the rows of the matrix 
$\bf S$, as $p$ {\it memorized patterns} embedded in the network
memory (it does not matter that not all the elements of the matrix $\bf S$ 
are equal $\{\pm 1\}$; see Note 1). Then the following meaningful 
interpretation of the problem can be suggested: {\it the network had to be 
learned by p-time showing of the standard 
$\vec\varepsilon (n)= (1,1,\ldots,1)$, but an error had crept into the 
learning process and in fact the network was learned with the help of its $p$ 
distorted copies; the value of the distortion $x$ was the same for all the
memorized patterns and every time only one coordinate had been distorted:} 
$$\vec s^{(l)}=(1,\ldots,1,\underbrace{1-x}_l,1,\ldots,1),\quad l=1,2,\ldots,
p.\eqno(4)$$ 

When $x$ is equal zero, the network is found to be learned by $p$ copies of 
the
standard $\vec\varepsilon (n)$. It is well-known that in this case the vector
$\vec\varepsilon (n)$ itself is the ground state and the functional has no
other local maxima. For continuity reasons, it is clear that the same 
situation remains for a sufficiently small distortion $x$. But when $x$ 
increases the ground state changes. For the problem (1)-(3) we
succeeded in obtaining the analytical description of the dependence of the
ground state on the value of the distortion parameter. 

{\bf Notations.} We denote by $\vec \varepsilon (k)$ the configuration vector 
which is collinear to the bisectrix of the principle orthant of the space 
$\rm R^k$. The vector, which after $p$ distortions 
generates the set of the memorized patterns $\vec s^{(l)}$, is called 
{\it the standard-vector}. Next, $n$ is the number of the spin variables, $p$ 
is the number of the memorized patterns and $q=n-p$ is the number of the 
nondistorted coordinates of the standard-vector. Configuration vectors are 
denoted by small Greek letters. We use small Latin letters to denote vectors 
whose coordinates are real. 

{\bf Note 1.} In the neural network theory the connection matrix obtained with
the help of Eq.(3) from $(p\times n)$-matrix $\bf S$ whose elements are equal
$\{\pm 1\}$ is called the Hebb matrix. If in Eq.(3) $(p\times n)$-matrix 
$\bf S$ 
is of {\it the general} type, the corresponding connection matrix will be 
called the matrix of {\it the Hebb type}. 

With regard to the set of the 
fixed points of the network an {\it arbitrary} symmetric connection matrix 
with zero diagonal elements is equivalent to a matrix of the Hebb type. 
Indeed, equality of the diagonal to zero guarantees the coincidence of the set
of the local maxima of the functional (1) with the
set of the network's fixed points. But the local maxima do not depend on the
diagonal elements of the connection matrix, so the lasts can be chosen 
whatever we like\cite{Lit3}. In particular, all the diagonal elements can
be taken so large that the connection matrix becomes a positive
definite one. And such a matrix can be already presented in the form
of the matrix product (3), where the elements of the related matrix $\bf S$ 
are not necessarily equal to $\{\pm 1\}$. In other words, with the help of 
the simple 
deformation of the diagonal a symmetric connection matrix turns into the Hebb
type matrix and as a result, the set of the local maxima of the
functional (1) does not change. This reasoning is correct for the
Hebb matrix too, since it is a symmetric one and  its  diagonal  elements 
are equal
zero. In such a way we ascertain the actuality of the Hebb type connection 
matrices for the Hopfield model (for details see \cite{Lit5}).

\vskip 1mm
${\bf 2^\circ. \mbox{ Basic model.}}$
Let us look for the local  maxima  among  configuration  vectors 
whose last coordinate is positive.

Since $q$ last columns of the matrix $\bf S$ are the same, the configuration
vector which is "under the suspicion" to provide an extremum is of the 
form\cite{Lit2}
$$\vec\sigma^*=(\underbrace{\sigma_1,\sigma_2,\ldots,\sigma_p}_{\vec\sigma'},
\underbrace{1,\ldots,1}_q),\eqno(5)$$
where {\it we denote by $\vec\sigma'$ the $p$-dimension part of the vector 
$\vec{\sigma}^{*}$, which is formed by its first $p$ coordinates.}
The direct calculations (or see \cite{Lit5}) show that  
$$F(\vec \sigma^*)\propto x^2 - 2x(q+p\cos w)\cos w +(q+p\cos w)^2,\eqno(6)$$
where
$$\cos w=\frac{\sum_{i=1}^p \sigma_i}p\eqno(7)$$
is the cosine of the angle between the vectors $\vec\sigma'$ and 
$\vec\varepsilon (p)$. Depending on the number of the coordinates of the 
vector
$\vec\sigma'$ whose value is "--1", $\cos w$ takes the values
$\cos w_k=1-2k/p$, where $k=0,1,\ldots,p$.
Consequently,  $2^p$ "suspicious-looking" vectors $\vec\sigma^*$ are 
grouped into the $p+1$ {\it classes } $\Sigma_k$:
the functional $F(\vec\sigma^*)$ has the same value $F_k(x)$ for all the 
vectors from the same class. {\it The classes $\Sigma_k$ are numerated
by the number $k$ of the negative coordinates which have the relevant
vectors $\vec\sigma^*$, and the number of the vectors in the $k$-class is 
equal
to $C_p^k$.}

To find the ground state under a given value of $x$, it is necessary to
determine the greatest of the values $F_0(x),F_1(x),\ldots,F_p(x)$. Under the
comparison the term $x^2$ can be omitted.
Therefore, to find out how the ground state depends on the parameter $x$, it 
is
necessary to examine the family of the straight lines
$$L_k(x)=(q+p\cos w_k)^2 - 2x(q+p\cos w_k)\cos w_k: \eqno(8)$$
in the region where the $L_k(x)$ majorizes all
the other straight lines, the ground state belongs to the class $\Sigma_k$ 
and is $C_p^k$ times degenerated. The analysis of the relative position
of the straight lines $L_k(x)$ gives\cite{Lit5}:

{\bf Theorem.} {\it As $x$ varies from $-\infty$ to $\infty$ the ground state 
in consecutive order belongs to the classes 
$\Sigma_0,\Sigma_1,\ldots,\Sigma_{k_{max}}$.
The jump of the ground state from the class $\Sigma_{k-1}$ into the class
$\Sigma_k$ occurs at the point $x_k$ of intersection of the straight lines
$L_{k-1}(x)$ and $L_k(x)$:
$$x_k=p\frac{n-(2k-1)}{n+p-2(2k-1)},\quad k=1,2,\ldots,k_{max}.$$
If $\frac{p-1}{n-1}<\frac13$, one after another all the $p$ rebuildings 
of the ground state take place according the
above scheme: $k_{max}=p$. And if $\frac{p-1}{n-1}>\frac13 $, the last
rebuilding is the one whose number is $k_{max}=\left[\frac{n+p+2}4\right]$.
The functional has no other local maxima.}

This theorem allows to solve a lot of practical problems. 

\begin{figure}[htb]
\begin{center}
\leavevmode
\epsfxsize = 16.2truecm
\epsfysize = 7.5truecm
\epsffile{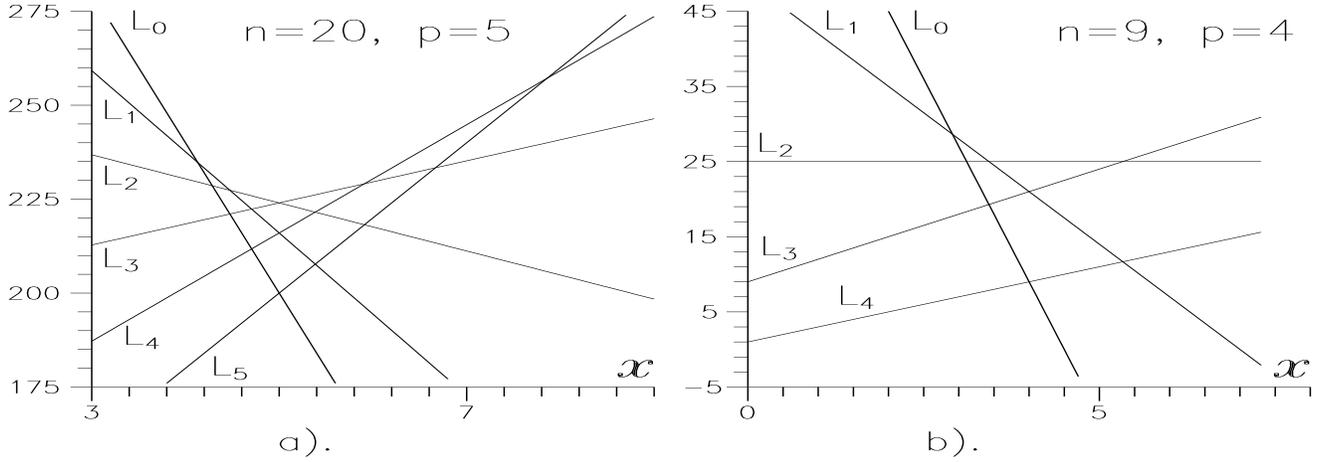}
\caption{ The typical behavior of the straight lines 
$L_k(x),\ k=0,1,\ldots,p$. The rebuildings of the ground state occurs at the
points $x_k$ of the intersection of the straight lines $L_{k-1}$ and $L_k$.
On the interval $(x_k,x_{k+1})$ the ground state is the one with the number 
$k$. When $x$ increases: a). all the rebuildings of the ground state occur
($k_{max}=5$), because $\frac{p-1}{n-1}<\frac13$; b). only 3 rebuildings of 
the ground state occur ($k_{max}=\left[\frac{n+p+2}4\right]=3$), because
$\frac{p-1}{n-1}>\frac13$.}
\end{center}
\end{figure}

In Fig.1 the typical examples of the relative position of the 
straight lines $L_k(x)$ are presented for the cases $\frac{p-1}{n-1}<\frac13$ 
(a) and $\frac{p-1}{n-1}>\frac13$ (b). When $x$ changes from $-\infty$ to 
$x_1$, the ground state is the standard-vector $\vec\varepsilon (n)$ (it 
exhausts the class $\Sigma_0$). In the point $x_1$ the ground state jumps from
the class $\Sigma_0$ to the class $\Sigma_1$ and becomes $p$ times 
degenerated.
When $x$ achieves the value $x_2$, the ground state jumps from the class 
$\Sigma_1$ to the class $\Sigma_2$ and becomes $C_p^2$ times degenerated, and
so on. As $x$ increases, the value of the functional {\it for the ground 
state}
at first monotonically decreases and then, after reaching the minimum value,
increases monotonically. For what follows let us note that
$k_{max}\ge [\frac{p+1}2]$, and $x_{k_{max}}\ge p$.

The case $p=n$ worth to be specially mentioned. Here all  the
jump points $x_k$ stick to one point
$$x'\equiv x_k =\frac{n}2,\quad k=1,2,3,\ldots,\left[\frac{n+1}2\right].
\eqno(9) $$
For any $x$ from the left of $x'$ the ground state is the standard-vector
$\vec\varepsilon (n)$, and for $x$ from the right of $x'$ the ground state
belongs to the class $\Sigma_{[\frac{n+1}2]}$ and is
$C_n^{[\frac{n+1}2]}$ times degenerated. 

The interval $x_1 < x < x_{k_{max}}$ will be called {\it 
the rebuilding region} of the ground state. This region is examined in details
in
\cite{Lit5}. Here we would like to mention only that its left boundary 
$x_1\ge\frac{p}2$ is the monotonically increasing function of $p$ as well as 
of
$n$. And also, when $p=const$ and $n\to\infty$ the rebuilding region tightens 
to the point 
$$x''=p.\eqno(10)$$
Here for  $x<x''$ the ground state is the standard-vector and for $x>x''$ 
the ground state belongs to the class 
$\Sigma_p$; again it is a nondegenerate one.  

{\bf Note 2.} The Theorem remains valid, if\cite{Lit5}: 

a). The memorized patterns (4) are normalized to unit to prevent their length
being dependent on the varying parameter $x$. As a result the maximum value of
the functional (1) {\it for the ground state} decreases monotonically as 
function of $x$. 

b). {\it An arbitrary} configuration vector $\vec{\alpha}=
(\alpha_1,\alpha_2,\ldots,\alpha_p,\alpha_{p+1},\ldots,\alpha_n)$ is 
used in place of the standard-vector $\vec\varepsilon (n)$. Then all the
results are formulated with respect to configuration vectors 
$\vec\sigma^*=(\alpha_1\sigma_1,\alpha_2\sigma_2,\ldots,\alpha_p\sigma_p,
\alpha_{p+1},\ldots,\alpha_n)$, and the elements of the connection matrix are
chanched: 
$$N_{ij}^{(\alpha)}=N_{ij}\alpha_i\alpha_j,\ i,j=1,2,\ldots,n.\eqno(11)$$

c). The first $p$ coordinates of the space $\rm R^n$ are subjected to a 
rotation. If in this connection the standard-vector does not change, all the
results of the "Basic model" are valid, though in this case the memorized
patterns are obtained from $\vec\varepsilon (n)$ by simultaneous distortion of
its $p$ coordinates! But if as a result of the rotation the standard-vector
turns into  
$$(u_1,u_2,\ldots,u_p,1,\ldots,1),\quad u_l\in{\rm R^1},\quad
\sum_{l=1}^p u_l^2=p,$$
the elements of the relevant connection matrix take the form
$$N_{ij}^{(U)}=N_{ij}u_iu_j,\ i,j=1,2,\ldots,n.\eqno(12)$$
Then, as in the "Basic model", the vectors $\vec\sigma^*$, which are "under the
suspicion" to provide an extremum, are grouped in the classes where the value
of the functional is constant. But now the vectors $\vec\sigma^*$ belong to the
same class if their p-dimensional parts are equidistant from the vector 
$\vec u=(u_1,u_2,\ldots,u_p)$. And just as before when $x$ increases, the 
ground state in consecutive order jumps from a class to the next one. Here the
Theorem remains valid, however a correction of the formulae is necessary. We
would like to note, that since the choice of the values $\{u_l\}_{l=1}^p$ is
completely in the researcher's hand, it is possible to construct the Hopfield
type networks with preassigned sets of fixed points.
An additional analysis is required to find out the limits of this method. 

We would like to mention that we succeeded in the generalization of the "Basic 
model" to the case when the linear term $h\sum_{i=1}^n\sigma_i$ was added to 
the 
functional $F(\vec\sigma)$ which had to be maximized. In physics models due to
such a term the external magnetic field can be taken into account. Now we 
prepare these results for publication.
\vskip 1mm 
${\bf 3^\circ.\mbox{ Interpretations.}}$
Let us in short discuss the results which are relative to the 
"Basic model".

{\bf Neural networks.} In this case the Theorem has to be
interpreted in the framework of the meaningful setting of the problem, which 
has been done above (see $1^\circ$): {\it the quality of 
"the truth" (the standard-vector) reconstruction by the network depends on the
distortion value $x$ during the learning stage and on the
length $p$ of the learning sequence}.

In agree with the common sense the error of the network increases with the
increase of the distortion $x$.

Also it is quite reasonable that the left boundary of the rebuilding region
$x_1$ is the increasing function of $p$ and $n$. Indeed, when $n$ and $x$ are
fixed, merely due to increase of the number of the memorized patterns $p$ the
value of $x_1$ can be forced to exceed $x$ (of course, if $x$ is not too  
large). As a result $x$ turns out to be left of $x_1$, i.e. in the region 
where
{\it the only} fixed point is the standard-vector. This conclusion is in 
agreement with the practical experience according which the greater the 
length of the learning sequence, the better the signal can be read
through noise. In the same way the increasing of $x_1$ with the increasing of
the number $n$ can be interpreted. 

When $p=const$ and $n\to \infty$ all the jump points $x_k$ 
stick to one point $x''=p$. In this case for $x<x''$ the ground state is the
vector which belongs to the class $\Sigma_0$:
$$\vec\varepsilon^{(+)} (n)=(\underbrace{1,1,\ldots,1}_p,1,\ldots,1),
\eqno(13)$$
and for $x>x''$ the ground state is the vector which belongs to the class 
$\Sigma_p$:
$$\vec\varepsilon^{(-)} (n)=(\underbrace{-1,-1,\ldots,-1}_p,1,\ldots,1)
\eqno(14)$$
(see Eq.(10)). As we see it, this result is a nontrivial one. In terms of 
the learning process, 
the distinct parts of the vectors $\vec\varepsilon^{(+)} (n)$ and 
$\vec\varepsilon^{(-)} (n)$ are two opposed statements. And the network 
"feels"
this. When the distortions $x$ is not very large (less than $x''$) the 
memorized
patterns $\vec s^{(l)}$ (4) are interpreted by the network as the distorted
copies of the vector $\vec\varepsilon^{(+)} (n)$ (13). But if during the 
learning 
stage the distortions exceed $x''$, the network interprets the memorized 
patterns $\vec s^{(l)}$ as the distorted copies of other standard-vector 
$\vec\varepsilon^{(-)} (n)$ (14). The last result is in agreement with the 
practical experience too: we interpret deviations in the image of a
standard as permissible ones {\it only till some threshold}. If only this 
threshold is exceeded, the  distorted  patterns  are  interpreted  as 
quite different
standard. (For details see \cite{Lit5}. In the same reference the very
interesting dependence of $k_{max}$ on the relation between $p$ and $n$ is
discussed.) 

{\bf The Ising model at {\sl T}=0.} The interpretation of this model in terms 
of the matrix
$\bf S$ is not known yet. Therefore here the obtained results are interpreted
starting from the form of the Hamiltonian $\bf N$ (3). Let's write it in the
block-matrix form: 
$$ \bf N\propto\left(\begin{array}{cc}
\bf A&\bf B\\
\bf B^T&\bf C\end{array}\right),$$
where the diagonal elements of the 
$(p\times p)$-matrix $\bf A$ and the $(q\times q)$-matrix $\bf C$ are equal
zero, and 
$$\left\{\begin{array}{ll}a_{ij}=1-2y,&i,j=1,2,\ldots,p,\quad i\ne j;\\ 
b_{ik}=1-y,&i=1,2,\ldots,p,\ k=1,2,\ldots,q;\\
c_{kl}=1,&k,l=1,2,\ldots,q,\quad k\ne l;\\
y=\frac{x}p.&\ \end{array}\right.$$
This matrix corresponds to a spin system with the infinitely large interaction
radius. The system consists of two subsystems, which are homogeneous with 
regard to the spin interaction. The interaction between the $p$ spins of the 
first subsystem is 
equal to $1-2y$; the interaction between the $q$ spins of the second subsystem 
is equal to $1$; the crossinteraction between the spins of each subsystems is
equal to $1-y$. When $p=n$ all the spins are interacting with each other in 
the same way. (We would like to remind that the connection matrix can be
generalized -- see Eqs. (11), (12).)

While $y<\frac12$ all the spins are interacting in the ferromagnetic way; when
$\frac12<y<1$, the interaction between the spins of the first subsystem 
becomes
of antiferromagnetic type, and when $1<y$ the crossinteraction is of
antiferromagnetic type too. The Theorem allows to trace how the ground state 
depends on the variation of the parameter $y$.

Let $p<n$. For $y\in (-\infty,\frac12)$ the ground state is the ferromagnetic 
one since $\frac12<y_1=\frac{x_1}p$, and for $x<x_1$ the ground state is the
standard-vector $\vec\varepsilon^{(+)} (n)$ (13). But it is interesting that
the ground state remains the ferromagnetic one even if $\frac12<y<y_1$, i.e.
when the
antiferromagnetic interactions already shown up in the system. In other words,
when $p<n$ there is {\it "a gap"} between the value of
the external parameter $y$ which corresponds to the destruction of the
ferromagnetic interaction and the value of this parameter which corresponds to
the destruction of the ferromagnetic ground state. Only after a "sufficient
amount" of the antiferromagnetic interactions is accumulated, the first jump 
of
the ground state occurs and it ceases to be the ferromagnetic one. Then after
another critical "portion" of the antiferromagnetic interaction is accumulated 
the
next jump of the ground state occurs (it happens when $y$ exceeds 
$y_2=\frac{x_2}p$), and so on. After the parameter $y$ reaches the value 
$y''=1=\frac{x''}p$, the crossinteraction becomes the antiferromagnetic one
too. The ground state continues "to jump" after that, since
$x_{k_{max}}\ge p$.

The energy $E=-\ F$ {\it of the ground state} as a function of the parameter
$y$ has breaks at the points $y_k=\frac{x_k}p$. It increases till $y\le y''=1$
and decreases when $y>y''$. However, if the memorized patterns are normalized
to unit, the energy of the ground state is a monotonically increasing function
of the external parameter.

It is natural to treat the case $p=const,\mbox{ }n\to\infty$ as the case of an
infinitely large sample with a few number of impurities. In this case all 
$y_k$ 
stick to the point $y''$ (see Eq.(10)). Depending only on the type of the
crossinteraction between the impurities and the rest of the sample, the ground 
state is either the ferromagnetic one (the vector $\vec\varepsilon^{(+)} (n)$ 
(13)), or the spins of the impurities are directed in an opposite way with
respect to the other spins of the sample (and the ground state is the vector
$\vec\varepsilon^{(-)} (n)$ (14)).

Finally, let's discuss the case $p=n$. Then all $y_k$ stick to the point 
$y'=\frac12$ (see Eq.(9)). Here the destruction of the ferromagnetic 
interaction
occurs simultaneously with the change of the ground state ( "the gap"
disappears). As long as the interaction of the spins is ferromagnetic 
($y<\frac12$), the ground state is ferromagnetic too. But when the interaction
of
the spins becomes antiferromagnetic ($y>\frac12$), the ground state turns out
to be $C_n^{[\frac{n+1}2]}$ times degenerated. From the right of $\frac12$ it 
is
natural to associate the state of the system with the spin glass phase.
\vskip 1 mm
{\bf Acknowledgments.}
The author is grateful to prof. A.A.Ezhov for helpful advices on the substance
of the work.

\end{document}